\documentclass[prl,a4paper,twocolumn,superscriptaddress,english]{revtex4-1}
\usepackage{babel}
\usepackage{xcolor}
\usepackage{graphicx}
\usepackage{amsmath}
\usepackage{amssymb}
\usepackage{wasysym}
\usepackage{hhline}
\newcommand{\bc}{\begin{center}}
\newcommand{\ec}{\end{center}}
\newcommand{\be}{\begin{equation}}
\newcommand{\ee}{\end{equation}}
\newcommand{\bea}{\begin{eqnarray}}
\newcommand{\eea}{\end{eqnarray}}
\newcommand{\dagga}{{\phantom{\dagger}}}

\definecolor{darkblue}{rgb}{0.1,0.2,0.6}
\definecolor{darkred}{rgb}{0.8,0.1,0.2}
\usepackage[colorlinks,citecolor=darkblue,linkcolor=darkred,urlcolor=darkblue]
{hyperref}
\usepackage[all]{hypcap} % fix figure link problem. Links jump to top of figure

\makeatletter
\adddialect\l@English\l@english
\makeatother

\newcommand{\rhosf}{\rho_\mathrm{SF}}

\definecolor{commentcolor_nl}{rgb}{0.1,0.2,0.6}
\definecolor{commentcolor_ed}{rgb}{0,0,1}
\definecolor{commentcolor_sc}{rgb}{1,0,0}
\definecolor{commentcolorD}{rgb}{1,0.1,.1}
\definecolor{todocolor}{rgb}{0.8,0.1,0.2}

\begin{document}
\title{Weak Versus Strong Disorder Superfluid-Bose Glass Transition in One Dimension}
\author{Elmer V. H. Doggen}
\affiliation{Laboratoire de Physique Th\'eorique, IRSAMC, Universit\'e de Toulouse, {CNRS, 31062 Toulouse, France}}
\author{Gabriel Lemari\'e}
\affiliation{Laboratoire de Physique Th\'eorique, IRSAMC, Universit\'e de Toulouse, {CNRS, 31062 Toulouse, France}}
\affiliation{Department of Physics, Sapienza University of Rome, P.le A. Moro 2, 00185 Rome, Italy}
\author{Sylvain Capponi}
\affiliation{Laboratoire de Physique Th\'eorique, IRSAMC, Universit\'e de Toulouse, {CNRS, 31062 Toulouse, France}}
\affiliation{Department of Physics, Boston University, 590 Commonwealth Avenue, Boston, Massachusetts 02215, USA}
\author{Nicolas Laflorencie}
\affiliation{Laboratoire de Physique Th\'eorique, IRSAMC, Universit\'e de Toulouse, {CNRS, 31062 Toulouse, France}}
\date{\today}

\begin{abstract}
Using large-scale simulations based on matrix product state and quantum Monte Carlo techniques, we study the superfluid to Bose glass-transition for one-dimensional attractive hard-core bosons at zero temperature, across the full regime from weak to strong disorder.
As a function of interaction and disorder strength, we identify a Berezinskii-Kosterlitz-Thouless critical line with two different regimes.
At small attraction where critical disorder is weak compared to the bandwidth, the critical Luttinger parameter $K_c$ takes its universal Giamarchi-Schulz value $K_{c}=3/2$. 
Conversely, a non-universal $K_c>3/2$ emerges for stronger attraction where weak-link physics is relevant.
In this strong disorder regime, the transition is characterized by self-similar power-law distributed weak links with a continuously varying characteristic exponent $\alpha$. 
\end{abstract}
\maketitle
\noindent{\it Introduction ---} The understanding of quantum interacting disordered systems represents an extraordinary challenge in condensed matter and quantum statistical physics. 
This has been recently discussed in the context of the transition to many-body localization (MBL)~\cite{Basko2006a, Nandkishore2015a}, for which a precise description is missing~\cite{Luitz2015a,Potter2015,Khemani2017}. 
Besides all these efforts made to understand such a localization transition at high energy, there are still fundamental open issues regarding the low-energy part of the problem. 

The most dramatic situation occurs in one dimension where quantum fluctuations plays a major role~\cite{Giamarchi2004}.
In the absence of disorder, interacting bosons may stabilize a quasi-ordered state at zero temperature, with a finite superfluid (SF) density and power-law decaying off-diagonal correlations $G_{ij}=\langle b^{\dagger}_i b^{\vphantom{\dagger}}_{j}\rangle\propto r^{-\frac{1}{2K}}$ at large $r \equiv |i-j|$, $K$ being the Luttinger liquid (LL) parameter~\cite{Giamarchi2004}. 
In the presence of disorder, such a SF ground state is generally expected to be unstable towards localization, thus forming the so-called Bose glass (BG) phase~\cite{Giamarchi1987a,Giamarchi1988a,Fisher1989a}: an inhomogeneous gapless compressible fluid with exponentially suppressed correlations. Disordered interacting chains can be experimentally realized using spin ladders~\cite{Klanjsek2008a,jeong_dichotomy_2016}, ultra-cold atoms~\cite{Lucioni2011a, DErrico2014a, Schreiber2015a} and arrays of disordered Josephson junctions \cite{Haviland2001a,Cedergren2017a,bard_superconductor_2017}. Broadly speaking, Luttinger liquid physics is expected to occur in various contexts~\cite{rmp}, e.g. quantum wires such as carbon nanotubes~\cite{nanotubes,nanotubes2}, fractional quantum Hall edge states~\cite{hall}, organic conductors~\cite{organic} and $^4$He confined in nanopores~\cite{He}.

In their seminal work, based on a LL description and renormalization group (RG) calculations at weak disorder, Giamarchi and Schulz (GS)~\cite{Giamarchi1987a,Giamarchi1988a}  have shown that a Berezinskii-Kosterlitz-Thouless (BKT) quantum phase transition should occur for a small but finite disorder strength, provided the clean LL parameter $K_0>3/2$. 
A recent two-loop calculation confirmed this scenario, with a universal jump at the BKT transition $K_c=3/2$~\cite{Ristivojevic2012a,Ristivojevic2014}. 
Below this value, SF is destroyed by quantum phase slips.

In order to go beyond this perturbatively weak disorder regime, an alternative picture has emerged
\cite{Altman2004a, Altman2008a, Altman2010a, Pielawa2013a},
based on a strong disorder scenario where the localization transition is predicted to occur for $K_c>3/2$ with a different mechanism driven by rare weak links that effectively cut the disordered chain. 
In this approach, whose asymptotic validity relies on growing disorder under RG,  a power-law tail for the distribution of Josephson couplings across weak links is expected
\be
P(J)\propto J^{\alpha},
\label{eq:PJ}
\ee
with $\alpha>0$ ($<0$) in the SF (BG), and $\alpha=0$ at the transition.
In a similar spirit, building on the idea of weak links, Pollet and co-workers~\cite{Pollet2013a,Pollet2014a,Yao2016a} suggested a third scenario. 
Connecting GS to strong disorder physics, they included  {a} renormalization of weak links {\it \`a la} Kane-Fisher~\cite{Kane1992a} and proposed a new ``scratched-XY" universality class to describe the SF-BG transition for $K_c>3/2$. 
At weak-link criticality, their approach predicts {that the critical LL parameter $K_c=\tilde{\alpha}/(\tilde{\alpha}-1)$ is governed by the \textit{non-universal bare} exponent $\tilde{\alpha}$ of {\it classical} Josephson couplings Eq.~\eqref{eq:PJ}}.
Despite several numerical studies~\cite{Bouzerar1994,Schmitteckert1998a,Hrahsheh2012a,Pielawa2013a,Yao2016a,Gerster2016}, the extension of the weak disorder regime remains unclear, and there is still no real consensus concerning  the scenario of the SF-BG transition  at strong disorder.

%%%%%%%%%%%%%%%%%%%%%
\begin{figure}[!b]
\includegraphics[width=\columnwidth]{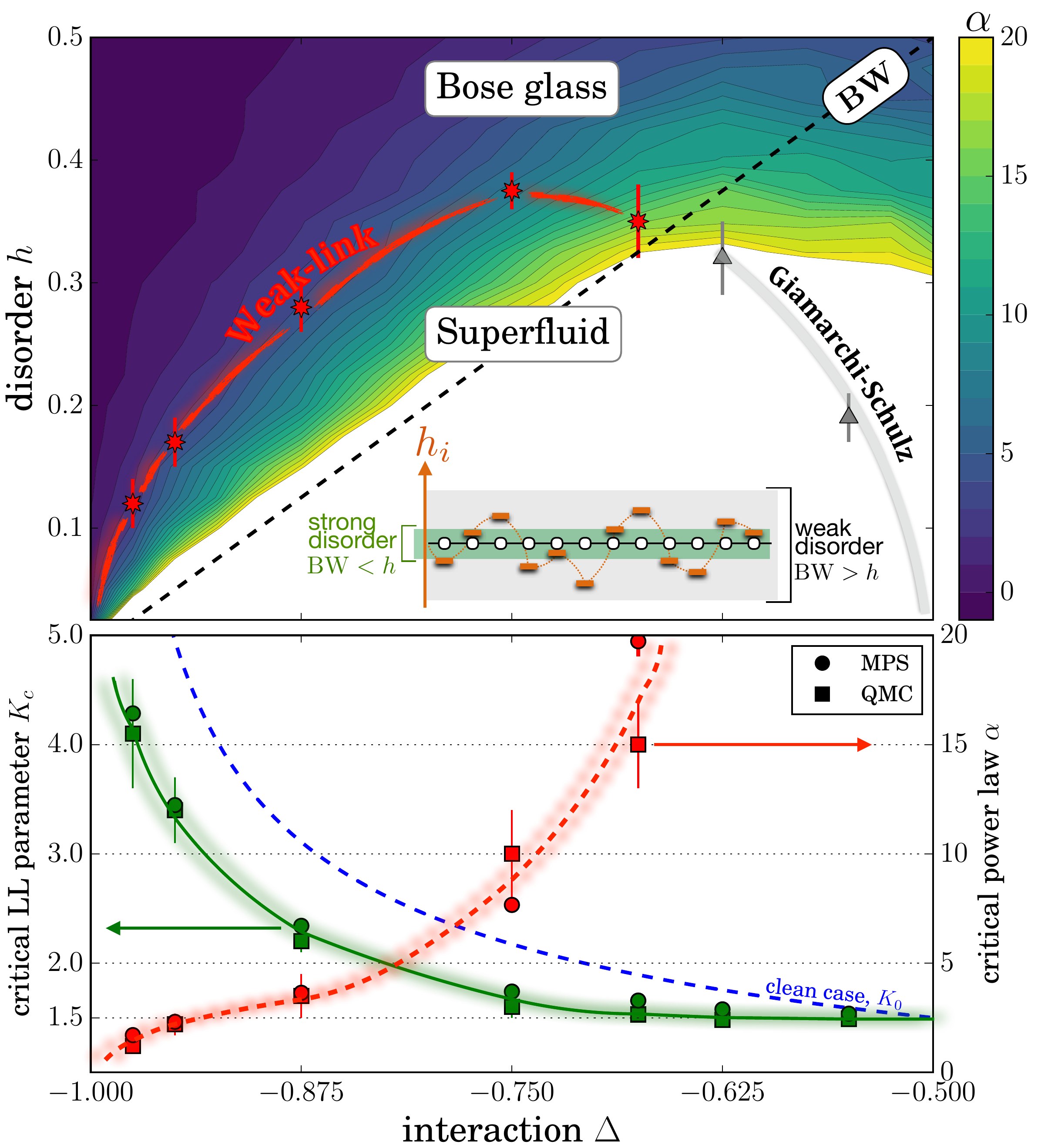}
\caption{\textbf{Top}: phase diagram, showing the SF and BG regimes. Red and gray symbols denote the SF-BG critical disorder $h_c$  for various values of the interaction $\Delta$. 
Color contours show the value of the power-law coefficient $\alpha$ from MPS simulations for $L=100$, with $-1\leq \alpha \leq 20$. $\alpha$ decreases for stronger attraction along the critical line, i.e.\ power law tails become fatter. 
Power-law fits fail in the white region below the bandwidth BW (black dashed line).  Inset: a random field configuration sketch is compared to BWs within weak (gray) {\it{vs.}} strong (green) disorder situations.
\textbf{Bottom}: Critical LL parameter $K_c$ (MPS and QMC estimates: green symbols, clean case value $K_0$: dashed blue line) and power-law coefficient $\alpha$ (red symbols) at the SF-BG transition for various values of the interaction $\Delta$. Red and green lines are guides to the eyes. $\alpha$ diverges at the crossover from the weak-link to GS regime. $L_{\rm MPS}=300$.
}
\label{fig:phase_diagram} 
\end{figure}

In this Letter, we elucidate this controversy using high-precision, large-scale numerical simulations of one-dimensional hard-core bosons in a random potential with nearest-neighbor attraction. 
Such a model is relevant to recent experiments using spin ladders in solid state systems \cite{Klanjsek2008a,jeong_dichotomy_2016}.

In the regime where the clean LL parameter $K_0>3/2$, we map out the entire phase diagram of model Eq.~\eqref{eq:H} as a function of interaction and disorder strengths (see Fig.~\ref{fig:phase_diagram}). 
We clearly identify a BKT {transition} line separating SF and BG regimes {by} a jump $K_c$ of the LL parameter. 
While $K_c=3/2$ for weak attraction and small disorder,
the situation is \emph{qualitatively} different when disorder exceeds the bandwidth, and weak links start to proliferate.
At criticality we clearly find $K_c>3/2$ and scale-invariant power-law distributions for the SF density and the off-diagonal correlations.
We strikingly observe a similar power-law tail exponent $\alpha$ for both quantities, which  validates a weak-link scenario.
Our results for $\alpha$ and $K_c$ along the critical line are different from the aforementioned predictions~\cite{Pielawa2013a,Yao2016a}.
Fig.~\ref{fig:phase_diagram} summarizes our results.\\

\noindent{\it Model and numerical approaches ---}  We study hard-core bosons on a chain, governed by the Hamiltonian
\be
{\cal{H}}=\sum_{i}\Bigl[-\frac{1}{2}\left(b_i^\dagger b_{i+1}^{\dagga} +b_i^\dagga b_{i+1}^\dagger\right)+\Delta n_i n_{i+1}-\mu_i n_i\Bigr].
\label{eq:H}
\ee
Hard-core bosons can be exactly mapped onto spin-1/2 using a Matsubara-Matsuda transformation, yielding the corresponding XXZ chain in a magnetic field
\be
{\cal{H}}=\sum_{i}\Bigl[-\left(S_i^x S_{i+1}^{x} +S_i^y S_{i+1}^{y}\right)+\Delta S_i^z S^{z}_{i+1}-h_i S_i^z\Bigr].
\label{eq:XXZ}
\ee
The random field{s} $h_i={\mu_i-\Delta}$ {are i.i.d.\ random variables box-distributed $\in [-h,h]$}.
{The half-bandwidth ${\rm{BW}}=1+\Delta$ is the relevant energy scale: in the absence of disorder, the XXZ chain is a polarized ferromagnet (an insulator in the bosonic language) for fields above ${\rm{BW}}$ and below $-{\rm{BW}}$. In the presence of disorder, if $h>\rm{BW}$ (above the dashed line in Fig.~\ref{fig:phase_diagram} top) there is a non-zero probability $P(r)\sim \left(1-{\rm{BW}}/h\right)^r$ to find a locally insulating region of length $r$, across which the SF response is exponentially suppressed, thus defining a weak link. In this strong disorder regime, it is already known that the {\it classical} local order parameter is power law-distributed \cite{Lemarie2013a}. There, rare disorder configurations may qualitatively change the physics. Conversely, $h < \rm{BW}$ defines the weak disorder regime. }

We have explored the ground state properties of model \eqref{eq:H} with two state-of-the-art  many-body numerical techniques. The matrix product state (MPS) formulation of the density-matrix renormalization group \cite{Schollwock2011a} {has been} used to capture the off-diagonal correlator $G_{ij}=\langle b^{\dagger}_i b^{\vphantom{\dagger}}_{j}\rangle$ for open chains up to $L=300$ sites. We {have} further use{d} the quantum Monte Carlo (QMC) algorithm in its stochastic series expansion version \cite{Syljuasen2002a} to estimate the SF density  $\rho_{\rm SF}$ and the compressibility $\kappa$ for closed rings up to $L=512$. In the latter case, ground state convergence is achieved at very low temperature using the $\beta$-doubling scheme \cite{Sandvik2002a}. Disorder averaging is performed over a large number of independent samples, from several hundred up to a few thousand.
Numerical details are discussed in the Supplementary Material {\cite{supmat}}.\\

%%%%%%%%%%%%%%%%%
\begin{figure}[!thb]
\includegraphics[width=\columnwidth]{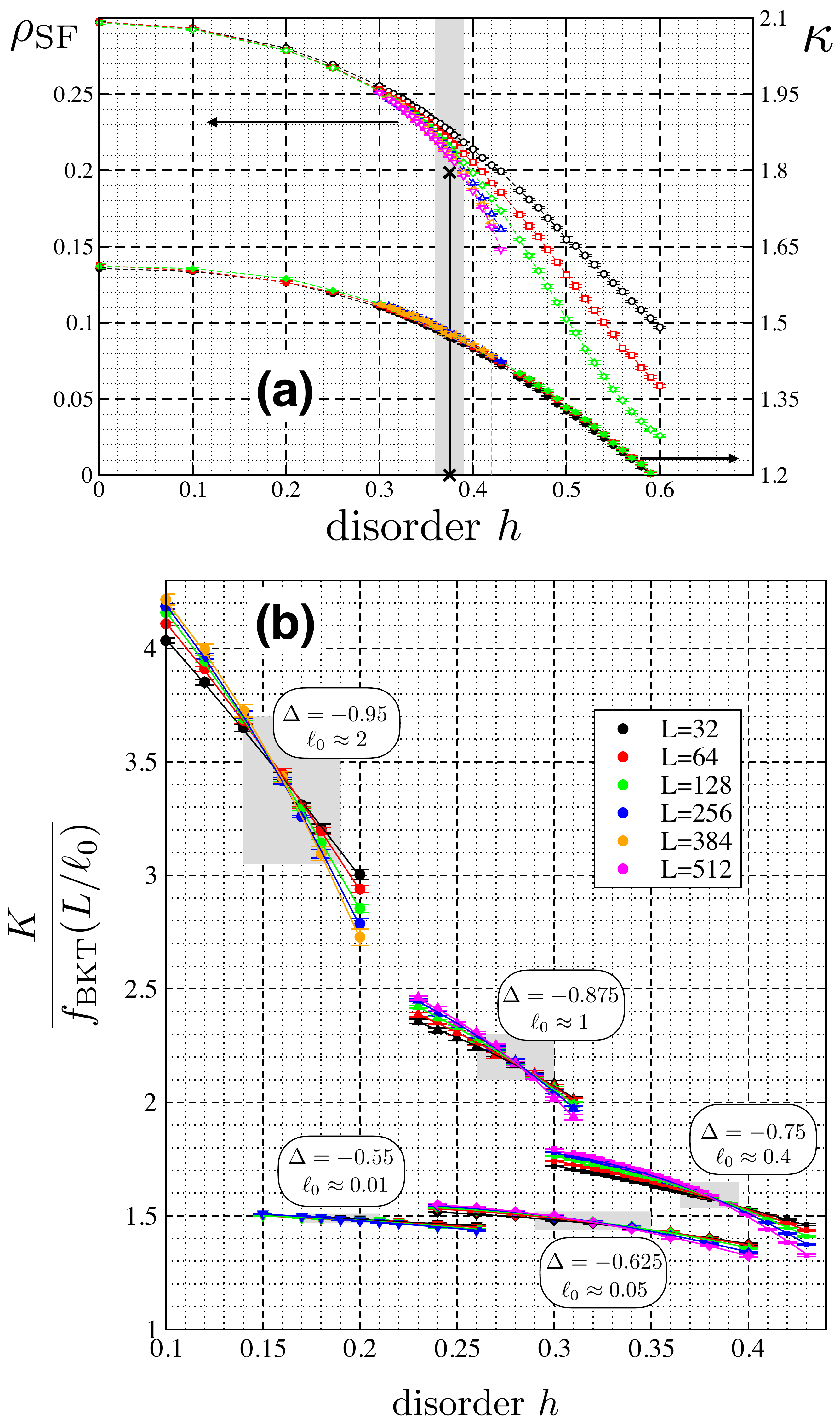}
\caption{QMC results for the  SF density $\rho_{\rm{SF}}$ (a) left, the compressibility $\kappa$ (a) right, and the LL parameter $K$ [Eq.~\eqref{eq:KLL}] (b). The $T=0$ converged data are averaged over $> 10^3$ independent random configurations. \textbf{(a)} Raw data for the SF density $\rho_{\rm SF}$ (left) and the compressibility $\kappa$ (right) across the SF - BG transition for $\Delta=-3/4$. $\rho_{\rm SF}$ jumps (black cross)  at the transition point $h_c=0.375(15)$ (gray region) while $\kappa$ remains smooth and featureless. \textbf{(b)} Scaling form of the LL parameter Eq.~\eqref{eq:BKT} plotted against disorder strength $h$ for various system sizes ($L=32,\ldots,512$) and various interaction strengths $\Delta$, as indicated on the plot. Crossings (gray area) are obtained using an $O(1)$ length scale $\ell_0$. }
\label{fig:BKT} 
\end{figure}
%%%%%%%%%%%%%%%%%

\noindent{\it Luttinger parameter and BKT transition ---} We first focus on the LL parameter $K$ which can be computed using MPS from the one-body density matrix \cite{Cazalilla2004a}
\be
G_{ij} \equiv \langle b^{\dagger}_i b^{\vphantom{\dagger}}_{j}\rangle\propto |i-j|^{-\frac{1}{2K}},
\label{eq:G}
\ee
and with QMC using the compressibility $\kappa$ and the SF density $\rho_{\rm SF}$ through the hydrodynamic relation~\cite{Giamarchi2004} 
\be
\label{eq:KLL}
K=\pi\sqrt{\kappa\rho_{\rm SF}}.
\ee
In the clean case $K_0(\Delta,h=0) = (\pi/2)(\pi-\arccos\Delta)^{-1}$ is exactly known from the Bethe ansatz solution of the XXZ chain~\cite{Yang1966a} (dashed blue line in the lower panel of Fig.~\ref{fig:phase_diagram}). We therefore focus on the attractive part of the phase diagram $\Delta\le -0.5$ where $K_0\ge 3/2$ such that a disorder-induced SF-BG transition is expected.

QMC results are displayed in Fig.~\ref{fig:BKT} panel (a) for both $\rho_{\rm SF}$ and $\kappa$ as a function of disorder strength $h$ for interaction $\Delta=-3/4$. In this case, a transition occurs (discussed below) at $h_c=0.375(15)$ between a SF regime at weak disorder and a BG above $h_c$. While the SF response vanishes in the BG at the thermodynamic limit $L\to \infty$, the compressibility remains smooth across the transition, with almost no size dependence. On the other hand $\rho_{\rm SF}(L)$, and therefore also $K(L)$ from Eq.~\eqref{eq:KLL}, displays quite strong finite size effects which are naturally expected for a BKT transition with  logarithmic corrections at criticality of the form $K(L)=K_c\times f_{\rm BKT}(L/\ell_0)$~\cite{Weber1988a,Laflorencie2001,Hsieh2013}, with $\ell_0$ a non-universal length scale:

\be
\label{eq:BKT}
f_{\rm BKT}(L/\ell_0)=1+\frac{1}{K_c\ln(L/\ell_0)}.
\ee
Assuming such a scaling form Eq.~\eqref{eq:BKT}~\cite{supmat}, the SF-BG transition is clearly detected~\cite{supmat}. In Fig.~\ref{fig:BKT} (b), the transitions for various values of $\Delta$ are characterized by different crossings obtained with a single free parameter, namely the $O(1)$ length scale $\ell_0$, as indicated on the plot.
From such an analysis, we can extract both $h_c$ and $K_c$ which are reported in Fig.~\ref{fig:phase_diagram}. The critical boundary $h_c$ {\it{vs.}} $\Delta$ shows two distinct behaviors: (i) {in the strong disorder regime $h_c>\,$BW, $K_c> 3/2$ is not universal ; (ii) in the weak disorder regime $h_c<\,$BW, the LL parameter takes the universal value} $K_c=3/2$ in agreement with GS scenario. While it is more difficult to locate the SF-BG critical point from MPS simulations{, the one-body density matrix $G$ \eqref{eq:G} has a decay controlled by $K$~\footnote{The LL parameter $K$ from QMC is a bulk quantity which displays logarithmic corrections of the form Eq.~\eqref{eq:BKT} at criticality whereas $K$ extracted from the decay of $G$ is an exponent whose finite size corrections are not known.} (see also \cite{supmat}), whose value is in very good agreement with QMC estimates}, as shown in Fig.~\ref{fig:phase_diagram} (green symbols, bottom).\\

\noindent{\it Critical distributions: emergence of weak-link physics ---} In order to further explore the qualitative differences between the two critical regimes, it is very instructive to go beyond disorder averages and investigate the {critical fluctuations.
Two quantities of significant interest are the distributions} of the one-body density matrix $G$, see Eq.~\eqref{eq:G},  (computed from MPS) and of the superfluid density $\rhosf$ (obtained from QMC). {Using MPS, we have considered the distribution $P_r[\ln \tilde{G}]$ at fixed $r \equiv |i-j|$ in the bulk of the system}~\footnote{$r \gg 1, |i-L/2| \ll L/2, |j-L/2| \ll L/2$}.
Here $\ln {\tilde{G}}=\ln G-{\overline{\ln G}}$, where ${\overline{\cdots}}$ denotes disorder averaging.
The subtraction of the typical value $\overline{\ln G}$ means that we are considering the fluctuations of $G$ on top of the characteristic decay $\propto r^{-1/2K}$.
These distributions are self-similar for different $r$~\cite{supmat}, which allows us to consider the averaged distribution $Q \equiv \overline{P_r}$ over $20$ different values of $r \in [20,40]$.
Using QMC, we study the distribution $P[\ln{\tilde{\rho}}_{\rm SF}]$ where $\ln{\tilde{\rho}}_{\rm SF}=\ln{{\rho}}_{\rm SF}-{\overline{\ln{{\rho}}_{\rm SF}}}$, which again explores the fluctuations {around} the typical SF response.

\begin{figure*}[t!]
 \includegraphics[width=2.0\columnwidth]{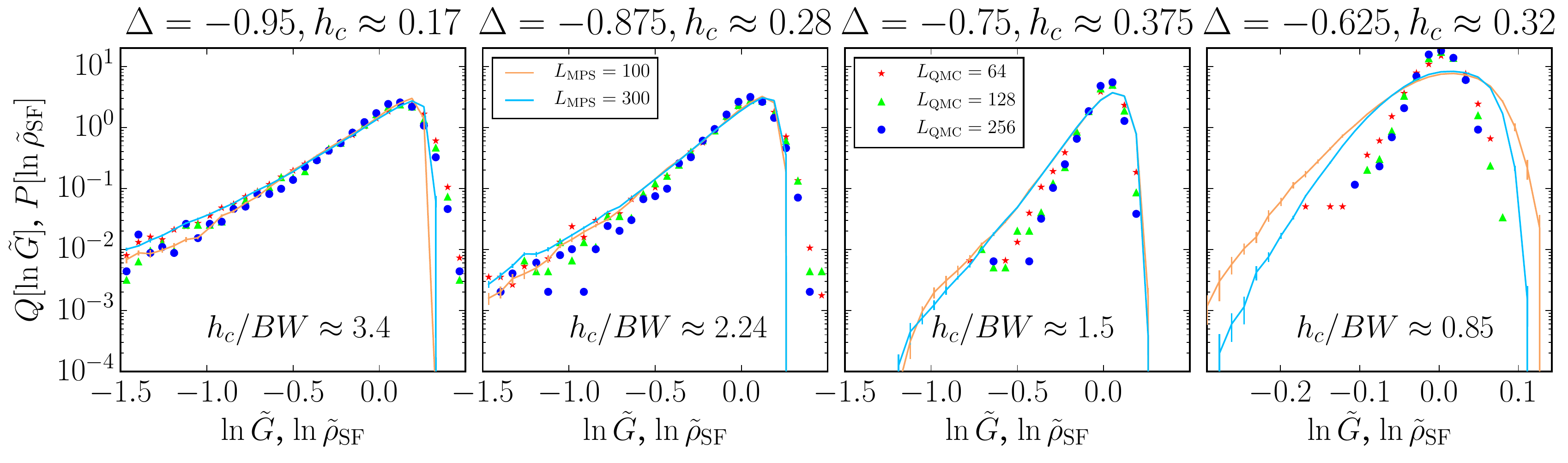}
 \label{fig:critdistr}
 \caption{Distributions of the correlator {$\ln \tilde{G} = \ln G - {\overline{\ln G}}$} (MPS, lines) and the superfluid stiffness {$\ln \tilde{\rho}_\mathrm{SF} =\ln{{\rho}}_{\rm SF}-{\overline{\ln{{\rho}}_{\rm SF}}}$ (QMC, symbols)} at the critical disorder $h_c$ for various values of $\Delta$. For $G$ we show $Q = \overline{P_r}[\ln \tilde{G}]$, where error bars are obtained from the dispersion of $P_r$ (see main text). Two regimes can be clearly distinguished: In the weak disorder regime (right panel) where $h_c<\,$BW, the distributions of $\ln \tilde{G}$ and $\ln \tilde{\rho}_\mathrm{SF}$ are distinct and self-averaging. Conversely, in the strong disorder regime $h_c>\,$BW, the distributions for both quantities overlap and do not depend on system size, indicating self-similarity. Moreover, they develop an exponential tail at low values Eq.~\eqref{eq:Powerlaw} controlled by the weak-link exponent $\alpha$, which decreases as $\Delta \rightarrow -1$ (very wide distributions, left panel) and diverges when approaching the GS regime (narrow distributions, second panel from the right). Note the differing $x$-axis scale on the last panel, indicating much narrower distributions.}
\end{figure*}

Representative results are shown in Fig.~\ref{fig:critdistr} for {four different} critical points. In the case of GS criticality (rightmost panel: $\Delta=-0.625, h_c=0.32(3), K_c=1.48(4)$), {the two} distributions are clearly distinct and their narrowing with increasing system size signals self-averaging. However, this is no longer the case {in the strong disorder regime}. Instead of self-averaging, we observe \emph{self-similar} distributions with exponential tails
\be\label{eq:Powerlaw}
\ln \left(P[\ln {\cal O}]\right)\sim (1+\alpha) \ln {\cal{O}},
\ee
{at low values of} ${\cal O}={\tilde{G}}$ or ${\tilde{\rho}}_{\rm SF}$, which corresponds to the aforementioned power-law distribution Eq.~\eqref{eq:PJ}. {Moreover, we strikingly observe a matching between the distribution of ${\tilde{G}}$ and ${\tilde{\rho}}_{\rm SF}$, which are \emph{independently} obtained}. The agreement improves when the critical disorder $h_c$ grows compared to the BW (see Fig.\ \ref{fig:critdistr}).
This result can be interpreted as a smoking gun for weak-link physics in the strong disorder regime, characterized by a well-defined scale-invariant exponent $\alpha$.

The weak link exponent $\alpha$ can be estimated from both $G$ (MPS) and $\rhosf$ (QMC) data using Eq.~\eqref{eq:PJ}. {Fig.~\ref{fig:phase_diagram} (red symbols, bottom) shows quantitative agreement, in particular at large $h_c/{\rm{BW}}$. When approaching the GS regime, weak-link physics becomes irrelevant, and $\alpha$ diverges. The color map in the phase diagram (Fig.~\ref{fig:phase_diagram} top) shows the evolution of $\alpha$ extracted from $G$ for $L=100$ chains. 
Note that in the BG regime this map has to be understood as a finite-size snapshot. Indeed, above the critical line self-averaging and self-similarity are lost and the distributions get broader with increasing system sizes {\cite{supmat}}. This is true throughout the entire BG regime, thus supporting the fact that despite the two different critical mechanisms, there is only a single BG phase in which  distributions get flatter, {with an} asymptotic behavior $\alpha\to -1$, in agreement with an infinite disorder scenario~\cite{Altman2010a}.

\vskip 0.45cm

\noindent{\it Conclusion. ---} Our numerical results convincingly show that the SF-BG transition for attractive hard-core bosons in a random potential follows two different routes, while BKT universality is maintained. 
As summarized in Fig.~\ref{fig:phase_diagram}, the GS scenario with a universal jump of the LL parameter $K_c=3/2$ occurs over a finite region of the phase diagram which corresponds to a critical disorder strength $h_c$ smaller than the bandwidth BW. 
In this weak disorder scenario, self-averaging is observed at the transition. 
On the other hand, for stronger attraction the bandwidth is smaller, and when $h>{\rm{BW}}$ rare insulating regions start to appear, yielding weak links across the system. 
In this strong disorder regime, the SF-BG transition occurs with a non-universal (interaction-dependent) jump of $K_c> 3/2$. 
Moreover, the critical fluctuations have a power-law distribution Eq.~\eqref{eq:PJ} with a single exponent $\alpha$ governing both the SF density $\rhosf$ and the one-body density matrix $G$.
Contrary to the strong disorder RG approach by Altman {\it{et al.}}~\cite{Altman2010a}, the weak-link exponent $\alpha$ continuously varies along the critical line, and diverges when approaching the GS regime. 
Our numerics also contrast with the ``scratched-XY'' universality of Pollet {\it et al.}~\cite{Yao2016a} where the LL parameter is predicted to obey $K_c={\rm{max}}(\frac{3}{2},{\frac{\tilde{\alpha}}{\tilde{\alpha}-1}})$, which disagrees with our results, assuming the bare exponent $\tilde{\alpha}=\alpha$~\footnote{We justify this through the observed scale invariance at criticality \cite{supmat}}.

Our conclusions clearly call for new analytical approaches {to describe} the strong disorder regime, as well as for the boundary between weak link and GS criticalities. 
As a final comment, one expects that the richness of the critical properties at play for the zero-temperature localization transition may influence the MBL physics at higher energy in the attractive regime we have studied. Moreover, weak link physics being a central concept for understanding rare regions at play for the MBL problem~\cite{agarwal_rare_2017,nahum_dynamics_2017}, we have provided a quantitative proof for such bottlenecks in the ground state. This should pave the way towards a better understanding of rare regions effects in quantum disordered systems.
Experimental verification of our results should be within reach using for instance solid state spin ladder systems or disordered cold atom setups.
\begin{acknowledgments}
\emph{Acknowledgments.---} We thank  C.\ Castellani, A.\ Polkovnikov, P.\ Simon and M.\ Rizzi for discussions, and E.\ Altman, G.\ Refael, Z.\ Yao, L.\ Pollet, N.\ V.\ Prokof'ev and B.\ V.\ Svistunov for useful comments on the manuscript. 
GL acknowledges an invited professorship at Sapienza University of Rome. SC thanks the Condensed Matter Theory Visitors Program at Boston University for financial support. 
The matrix product state computations in this work have been carried out using Open Source Matrix Product States ({\small{OSMPS}}) \cite{OpenMPS2012,OpenMPS2017}.
We thank {\small{CALMIP}}  and {\small{GENCI-IDRIS}} (Grant 2016-050225) for providing computational resources and M.\ Dupont and D.\ Jaschke for technical assistance.
This work is supported by the French ANR program BOLODISS (Grant ANR-14-CE32-0018) and  by Programme Investissements d'Avenir under the program ANR-11-IDEX-0002-02, reference ANR-10-LABX-0037-NEXT.
Furthermore, we acknowledge the use of {\small{LAPACK}} \cite{Lapack1999}, {\small{GNU}} Parallel \cite{gnuparallel} and Matplotlib \cite{Matplotlib}.
\end{acknowledgments}

\appendix

\section{Supplementary material}

\renewcommand{\theequation}{S\arabic{equation}}
\renewcommand{\thefigure}{S\arabic{figure}}
\setcounter{figure}{0}
\setcounter{equation}{0}

\section{Technical details (MPS)}
In the following, we will outline the numerical approach for the matrix product state (MPS) simulations.
As a reminder, we seek to obtain the ground state at half filling of the following Hamiltonian:
\begin{equation}
 \mathcal{H} = \sum_{i} \Bigl[ -\frac{1}{2} b_i^\dagger b_{i+1} + \mathrm{H.c.} + \Delta n_i n_{i+1} + \mu_i n_i \Bigr], \label{eq:DMRGHam}
\end{equation}
where $b_i^{(\dagger)}$ destroys (creates) a hard-core boson at site $i$ on a chain of length $L$, $n_i \equiv b_i^\dagger b_i$ and $\Delta$ is the strength of nearest-neighbour interactions.
The on-site disorder is given by $\mu_i \in [-h,h]$ with a uniform distribution of the disorder. MPS calculations are performed in the (conserved) sector $\sum_i \langle n_i\rangle =L/2$, such that exact half-filling is imposed.
We use Open Source Matrix Product States (OSMPS) \cite{OpenMPS2012,OpenMPS2017}.
Since the ground state algorithm uses a variational approach starting from an initial educated guess, convergence to the ground state is not formally guaranteed.
In particular, there is the risk of the algorithm getting ``stuck'' in a metastable state, which is known to affect disordered systems that may have several local minima of the energy.
To alleviate this risk, we perform as many DMRG sweeps as necessary up to a maximum of $10$ (for a clean system, usually only a few sweeps are necessary), using a maximum bond dimension of $\chi = 300$ until the following criterion is met:
\begin{equation}
 \langle \mathcal{H}^2 - \langle \mathcal{H} \rangle^2 \rangle < \epsilon L,
\end{equation}
where the tolerance per site $\epsilon = 10^{-10}$.
Numerically, our results are obtained by finding the ground state of \eqref{eq:DMRGHam} using these parameters for a total of $R$ realizations using an embarrassingly parallel approach and storing the full $G_{ij} \equiv \langle b_i^\dagger b_j \rangle$ for each realization.

\begin{figure}[!htb]
 \includegraphics[width=\columnwidth]{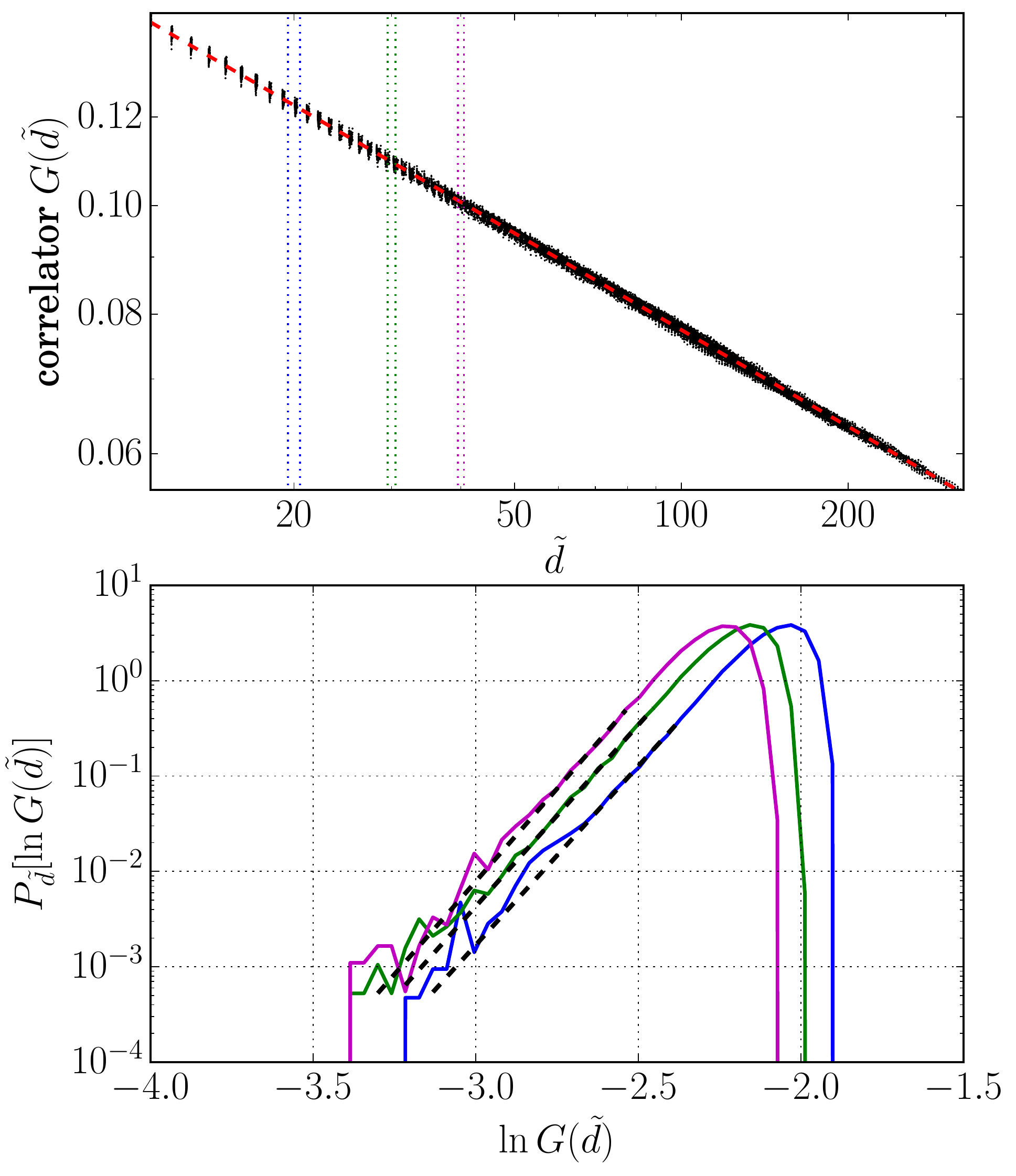}
 \label{fig:dmrgK}
 \caption{\textbf{Top} (note the double log scale): Determination of the Luttinger liquid parameter $K$ using matrix product state simulations for a system of size $L=300$ for $\Delta = -3/4, h=0.375$, close to the critical disorder $h_c$. Each black dot represents the typical value of $G_{ij}$ over all $R = 280$ realizations, plotted as a function of the rescaled distance $\tilde{d}(i,j)$. The red dashed line represents a least-squares fit to a function decaying as a power law, yielding $K \approx 1.74(3)$ where the error is estimated through a bootstrapping procedure. 
 \textbf{Bottom} (note the log scale): probability density function of $\ln G_{ij}$ in small windows $\tilde{d} \in [19.5,20.5]$ (blue), $\in [29.5, 30.5]$ (green), $\in [39.5,40.5]$ (magenta). A fat power law-tail, indicating an elevated probability of finding weak links, is visible for small values of $G$. The distribution consists of all values of $G$ inside the ``slices'' shown in the top panel (dotted lines), i.e.\ $R=280$ values per each black dot. Aside from the shift due to the decay $\propto \tilde{d}^{-1/2K}$, the distributions are self-similar to an excellent approximation. Black dashed lines indicate an exponential fit, corresponding to a power law in $G$. Averaging many such power law fits over different slices leads to the estimate $\alpha = 7.7(3)$, where the error is determined through the dispersion of the data.}
\end{figure}

\emph{Determining the Luttinger liquid parameter $K$.---} We describe in detail the procedure to determine the decay of the single-particle density matrix or correlator $G$:
 \begin{equation}
 G_{ij} = \langle b_i^\dagger b_j \rangle, \label{eq:densmatrix}
\end{equation}
which in the superfluid (SF) regime obeys an algebraic decay as $\propto r^{-1/2K}$ with $r \equiv |i-j|$.
In a finite system with open (infinite wall) boundary conditions, $G$ will vanish at the boundary of the system.
To correct for this, we use a method based on conformal field theory \cite{Cazalilla2004a}.
According to this theory, the value of $G$ in a finite system is described to leading order by:
\begin{subequations}
\label{eq:caza}
\begin{align}
 G_{ij} = \rho_0 B_0 \Big[ \tilde{d}(i,j) \Big]^{-1/2K}, \label{eq:caza1}\\  
 \tilde{d}(i,j) \equiv \frac{d(i+j|2L)d(i-j|2L)}{\rho_0^{-1} \sqrt{d(2i|2L)d(2j|2L)}}, \label{eq:caza2}
\end{align}
\end{subequations}
where $d$ is the ``chord function:''
\begin{equation}
 d(i|L) = \frac{L}{\pi}|\sin(\pi i/L)|.
\end{equation}
$B_0$ is a non-universal constant and $\rho_0$ is the average density (these constants are not important for our purposes).
The numerical procedure to determine $K$ now consists of computing $G_{ij}^{(\nu)}$ for each realization $\nu = 1 \ldots R$ and all $i,j \in [1,L]$.
Next, we compute both the typical value $G_{ij,\mathrm{typ}} \equiv \exp(\frac{1}{R} \sum_\nu \ln G_{ij}^{(\nu)})$ and the average $G_{ij,\mathrm{avg}} \equiv \frac{1}{R} \sum_\nu G_{ij}^{(\nu)}$.
$K$ can now be determined using the Levenberg-Marquardt least-squares fitting algorithm using either the typical or average $G$.

Although the approach \eqref{eq:caza} captures the most important finite-size corrections, it is still more accurate in the bulk than near the edges of the system or for very short-range corrections.
Hence, we consider only $r > 10$ and discard values of $G$ with $i,j$ close to the edge of the system, specifically we consider only $|i-L/2| < L/3$ and $|j-L/2| < L/3$.
In the clean case, the exact value for $K$ is known from the Bethe Ansatz solution \cite{Yang1966a}:
\begin{equation}
 K = \frac{\pi}{2(\pi-\arccos\Delta)}.
\end{equation}
Benchmarking to the non-interacting case at $\Delta = -0.75$ gives a numerical result that is approximately $0.1\%$ off from this exact result.
An example of this procedure to determine $K$ is shown in Figure \ref{fig:dmrgK} for $\Delta = -3/4$ and $h = 0.375$, at the transition from SF to BG.
For this value and system sizes $L=100$, $L=200$ and $L=300$ we obtain $K = 1.76(5)$, $K = 1.74(3)$ and $K = 1.74(3)$ using $761$, $368$ and $280$ realizations respectively.
Although no reliable finite-size scaling procedure is known to extrapolate to $L \rightarrow \infty$, these numerical results (consistent with results for other values of $\Delta,h$) suggest that the finite-size corrections converge to the thermodynamic result more quickly than QMC results using the superfluid density and compressibility, so that the finite-size result for $L = 300$ is fairly close to the QMC estimate in the thermodynamic limit (see Fig.\ 1 of the main manuscript). This also opens up avenues for accurate experimental measurements of $K$ using finite systems yielding only a modest overestimate of $K$, provided an accurate method to measure the correlations $\langle b_i^\dagger b_j \rangle$ is used.

We briefly remark on the difference between using the typical and average $G$ for determining $K$. Numerically, we find that in the SF regime they yield almost identical results. However, in the BG regime they start to deviate and moreover the estimate for $K$ develops a strong system size dependence. The point at which the typical $K_\mathrm{typ}$ and average $K_\mathrm{avg}$ estimates for $K$ start to deviate from each other yields a qualitative criterion for the SF-BG transition which is in good agreement with estimates of the critical disorder from QMC. However, an accurate determination of the critical disorder would require detailed knowledge of the scaling of $|K_\mathrm{typ}-K_\mathrm{avg}|$ close to the transition, which is \emph{a priori} unknown.

As a final remark, we stress that the decay of the correlator as shown in Fig.\ \ref{fig:dmrgK} should not be confused with the distribution of $P_r[\ln G]$ as discussed in the main text. The distributions in the latter case are obtained by looking at a \emph{fixed} $r = |i-j|$ (or fixed $\tilde{d}$, which yields similar distributions for sufficiently small variations of $r$ or $\tilde{d}$) and then collecting all values of $G$ (not just the typical value) within a narrow window around $r$ or $\tilde{d}$. Thus, one takes a small ``slice'' of values in Fig.\ \ref{fig:dmrgK} and then computes the probability density function $P_r[\ln G]$ using $R$ values per each $G_{ij}$ inside the slice. The total distribution $Q = \sum_r \overline{P_r}$ is then obtained by summing and averaging several such contributions in consecutive slices, while subtracting the typical value to correct for the influence of the decay $\propto r^{-1/2K}$ so that consecutive $P_r$'s are self-similar. We illustrate this procedure in Fig. \ref{fig:dmrgK} (bottom panel). After all these steps, only a modest number of realizations is required for very accurate distributions (see Fig. 3 of the main paper).

\begin{figure*}[!htb]
 \includegraphics[width=1.7\columnwidth]{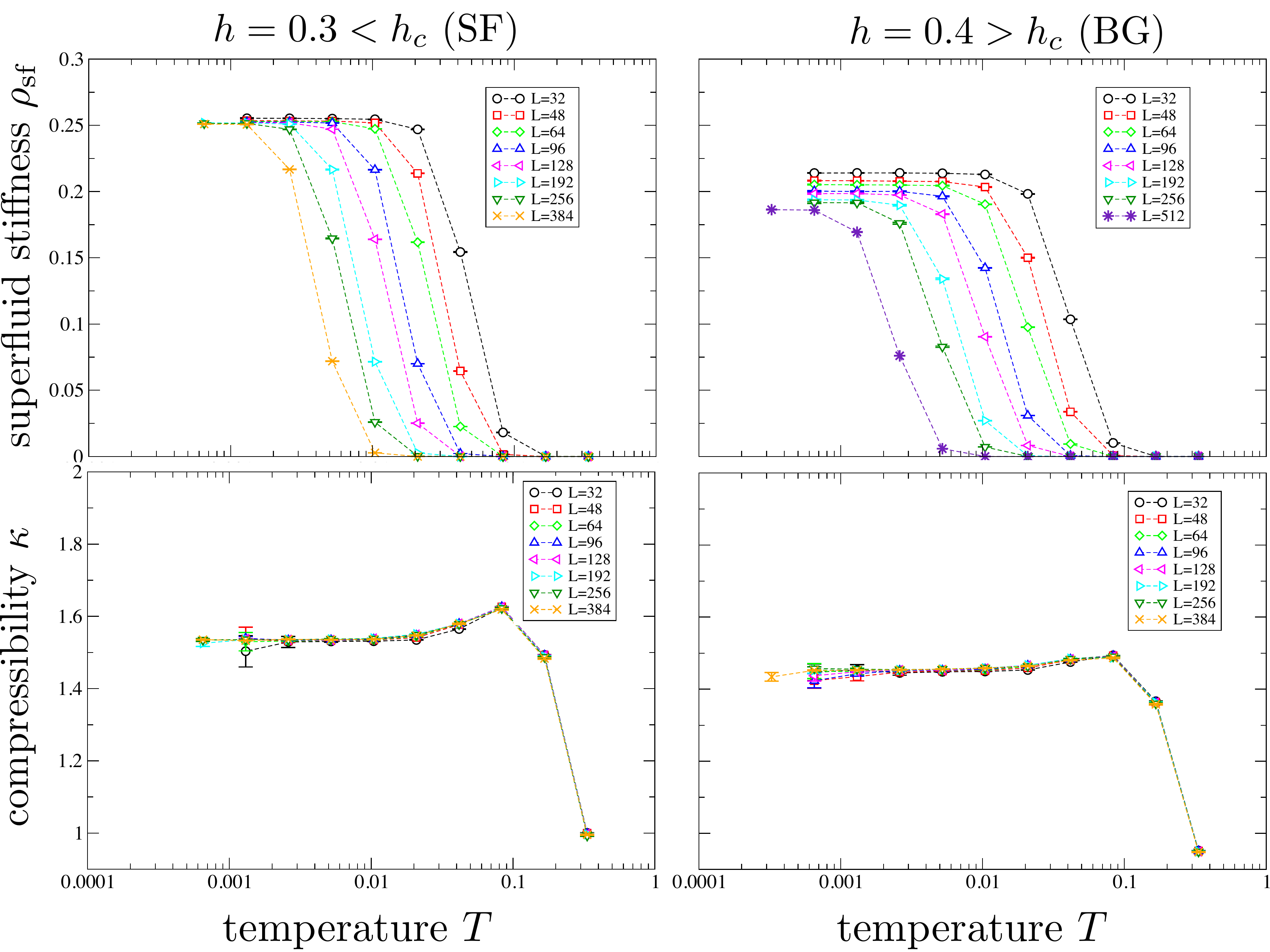}
 \label{fig:beta}
 \caption{Ground state convergence using the $\beta$-doubling scheme. Quantum Monte Carlo results for the superfluid stiffness $\rho_{\rm sf}$ (top) and the compressibility $\kappa$ (bottom) are shown for $\Delta=-0.75$ and two disorder strengths $h=0.3$ (SF regime, left) and $h=0.4$ (BG regime, right). Here the data have been averaged over $\sim 2000$ disordered samples.}
\end{figure*}

\section{Technical details (QMC)}
The Quantum Monte Carlo (QMC) simulations have been performed using the Stochastic Series Expansion (SSE) method~\cite{Syljuasen2002a}. While this is a rather standard technique for non-frustrated quantum spin models in the absence of disorder, in the presence of randomness one has to pay some attention regarding several issues, as discussed in previous works~\cite{Sandvik2002a,Laflorencie2004}:
\begin{itemize}
\item[1)] Equilibration and thermalization.---
In order to achieve a fast equilibration, the $\beta$-doubling scheme~\cite{Sandvik2002a} is used, cooling down the system following $\beta=3,\,6,\,12,\cdots,3*2^p$, with $p_{\rm max}=10-11$, depending on the system size. At each temperature step, $N_{\rm eq}=10^3$ MC steps are used.
\item[2)] Ground state convergence.---
The convergence to $T=0$ properties is also achieved using the $\beta$-doubling procedure, as exemplified below in Fig.~\ref{fig:beta} for $\Delta=-3/4$ and various disorder strengths.

\item[3)] Monte Carlo {\it{vs.}} sample-to-sample fluctuations.--- In the strong disorder regime, using $N_{\rm mes}=10^3$ MC steps of measurement is enough since the distributions displays fat tails. On the other hand, in order to correctly sample such power-law tails, several thousands of independent disordered samples have been used. In the weak disorder regime, where distributions are much narrower, we have used $10^4$ and up to $10^5$ MC steps in order to correctly capture self-averaging.
\end{itemize}
We also note that simulations have been performed in the grand-canonical ensemble, where particle number can fluctuate. Regarding the random field configurations, we did {\it{not}} impose any constraint such as $\sum_i h_i=0$ over a sample. Instead, each sample has an independent configuration taken in a uniform box distribution in the range $[-h,h]$. We do not expect the distinction between canonical {\it{vs.}} microcanonical ensemble for the disorder~\cite{Monthus2004} to change the critical properties, as demonstrated for a related study in two dimensions~\cite{Alvarez2015,Ng2015}.

\begin{figure}[!htb]
 \includegraphics[width=\columnwidth]{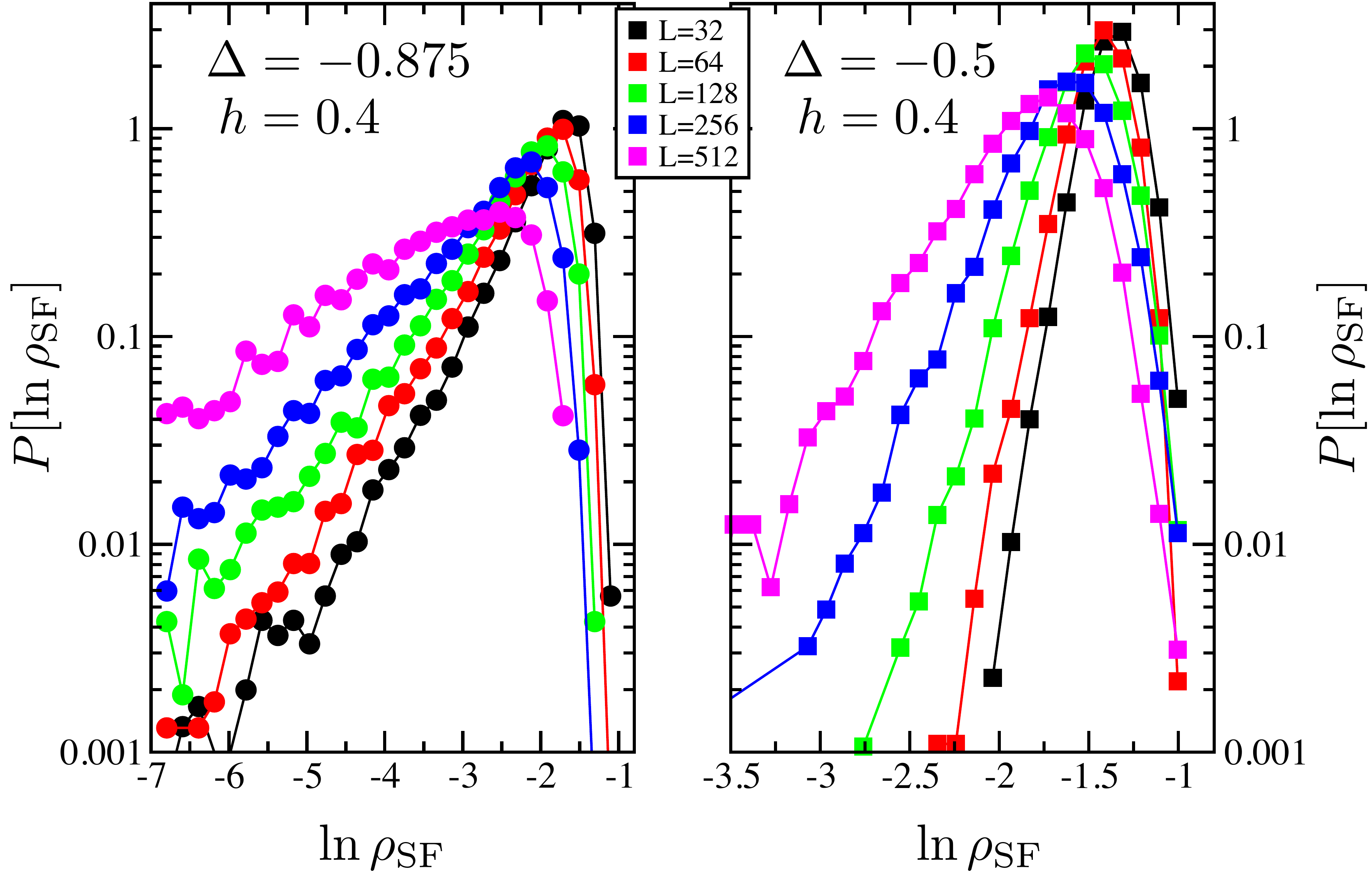}
 \label{fig:BG}
 \caption{Quantum Monte Carlo results for the distribution of superfluid densities in the Bose glass regime, either above (left) of below (right) the BW line. In both cases, broadening with system size is clearly observed, {\it i.e.}\ an absence of self-averaging.}
\end{figure}

\begin{figure*}[!htb]
 \includegraphics[width=1.7\columnwidth]{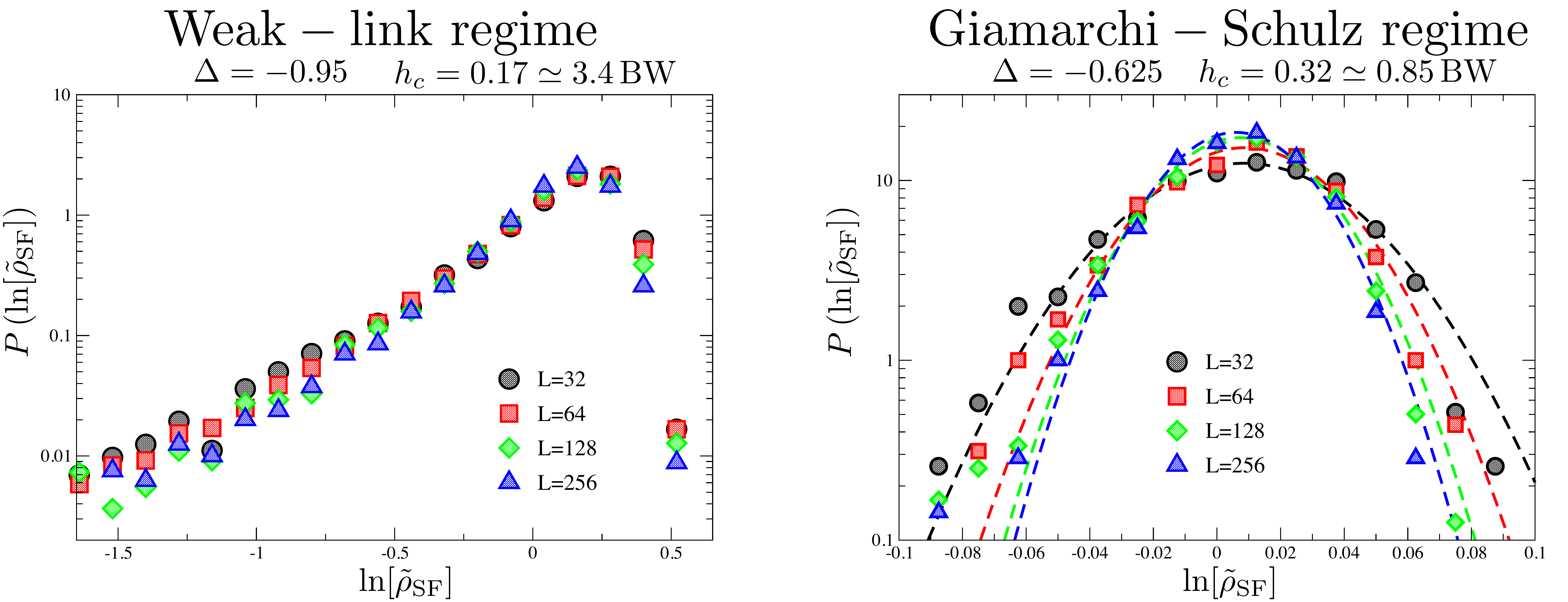}
 \label{fig:rhosf_crit}
 \caption{Critical distributions of the superfluid density $\tilde{\rho}_\mathrm{SF}$ deep in the weak-link regime and in the weak disorder (Giamarchi-Schulz) regime. In the weak-link regime, no system size $L$-dependence is observed, while the distributions are self-averaging (shrinking with increasing $L$) in the weak disorder regime where dashed lines are gaussian fits. Note the differing $x$-axes.}
\end{figure*}

%\newpage
\section{Broadening of the  superfluid stiffness distribution in the Bose glass regime}
As discussed in the main text, the weak-link distributions in the Bose glass regime (at $h>h_c$) get broader with increasing system size. This is shown in Figure \ref{fig:BG} where QMC results for $P(\ln\rho_{\rm sf})$ are displayed for two representative values of interaction $\Delta=-0.5$ and $-0.875$ at the same disorder strength $h=0.4$. Here we see that both cases are qualitatively similar, despite the fact that\\
(i) at $\Delta=-0.5$, $h/{\rm{BW}}=0.8$, corresponding to ``weak" disorder;\\
(ii) at $\Delta=-0.875$, $h/{\rm{BW}}=3.2$, which is far above the BW line, {\it{i.e.}} in the ``strong" disorder regime.

From the behavior of these distributions, qualitatively similar to what was observed for the Bose glass regime in two dimensions~\cite{Alvarez2015}, there is no distinction between ``weak" and ``strong" disorder regime, and we expect only a single Bose glass phase with $\alpha\to -1$ when $L\to\infty$.

\section{Critical distributions and power law $\alpha$}

As discussed in the main manuscript, the distributions of the superfluid density $\tilde{\rho}_\mathrm{SF}$ and correlator $Q[\ln \tilde{G}]$ show two qualitatively different regimes: a self-similar regime if $h_c > \rm BW$, and a self-averaging regime for $h_c <\rm  BW$. 
In Figures \ref{fig:rhosf_crit}-\ref{fig:mps_crit} we show two characteristic examples of critical distributions for QMC and MPS data separately. Since $\rhosf$ and $G_{ij}$ are entirely different quantities, there is no \emph{a priori} reason to expect the distributions to be similar. That they are, nonetheless, very similar at the transition in the weak-link regime shows the crucial importance of the power-law tails associated with weak links in this regime. Indeed, in the weak disorder regime $h_c < \rm BW$ the distributions are very different although still self-averaging for both QMC and MPS data.

\begin{figure}[!htb]
 \includegraphics[width=1\columnwidth]{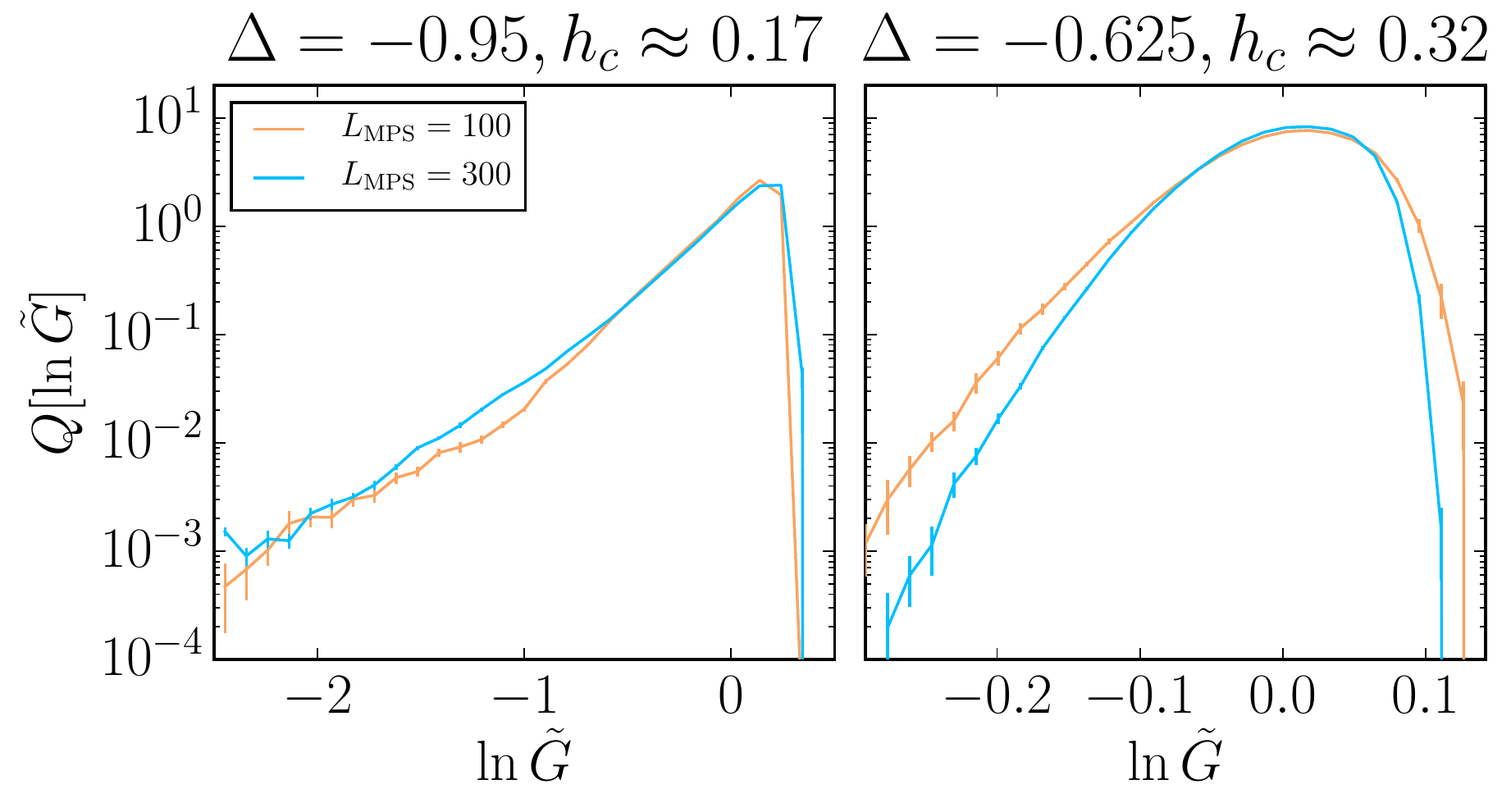}
 \label{fig:mps_crit}
 \caption{As Figure \ref{fig:rhosf_crit}, but for $Q[\ln \tilde{G}]$ using MPS data.}
\end{figure}

It is also instructive to consider the behavior of the power law $\alpha$ near the superfluid-Bose glass transition, as shown in Figure \ref{fig:qmc_alpha}. 
A qualitative difference between the distributions of $\rhosf$ and $G$ is that the distributions for $\rhosf$ are self-similar only exactly at the transition, while for MPS data they remain self-similar in the superfluid regime. This means that the value of $\alpha$ shown in the color contour plot of Fig.\ 1 of the main manuscript should be a good approximation of the value of the power law in the superfluid regime in the thermodynamic limit. While we do not observe a change in the power law tail as a function of $L$ for $Q[\ln G]$ in the BG regime, we do observe a broadening of the peak of the distribution (not shown), which should eventually remove the power law in the limit $L \rightarrow \infty$. On the other hand, the data for $\rhosf$ shows this trend already for modest $L$ (see Figure \ref{fig:qmc_alpha}).\\
\\

\begin{figure}[!htb]
 \includegraphics[width=\columnwidth]{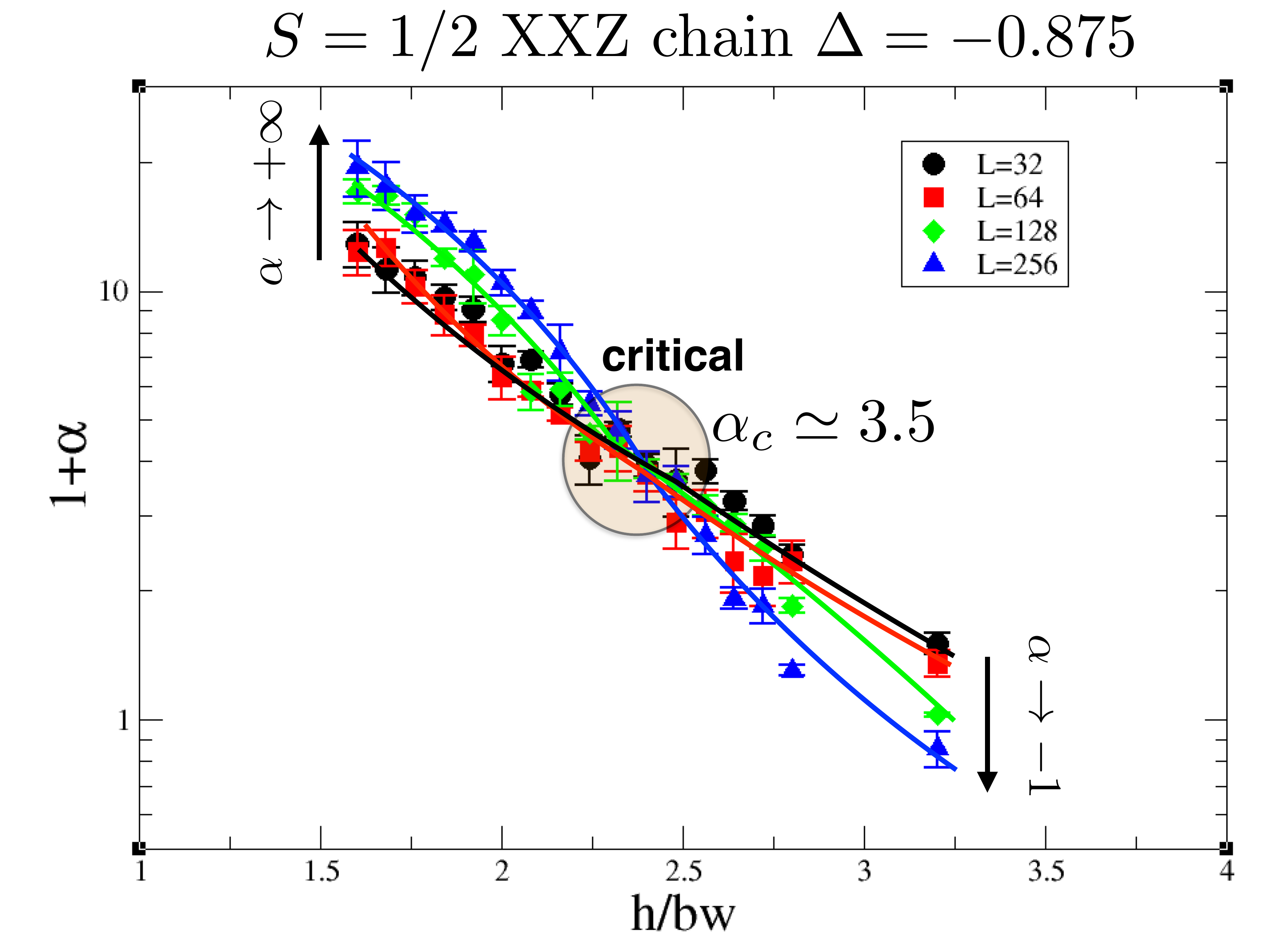}
 \label{fig:qmc_alpha}
 \caption{Power law coefficient $\alpha$ measuring the decay of the distribution of the superfluid density $\rhosf$ at small values, as a function of the ratio of disorder $h$ and half-bandwidth $\rm BW$. In this case $\Delta=-0.875$, self-similarity occurs at $h_c\sim 2.3\rm BW$ with a critical power-law $\alpha_c\simeq 3.5$.}
\end{figure}

\begin{figure}[!htb]
 \includegraphics[width=\columnwidth]{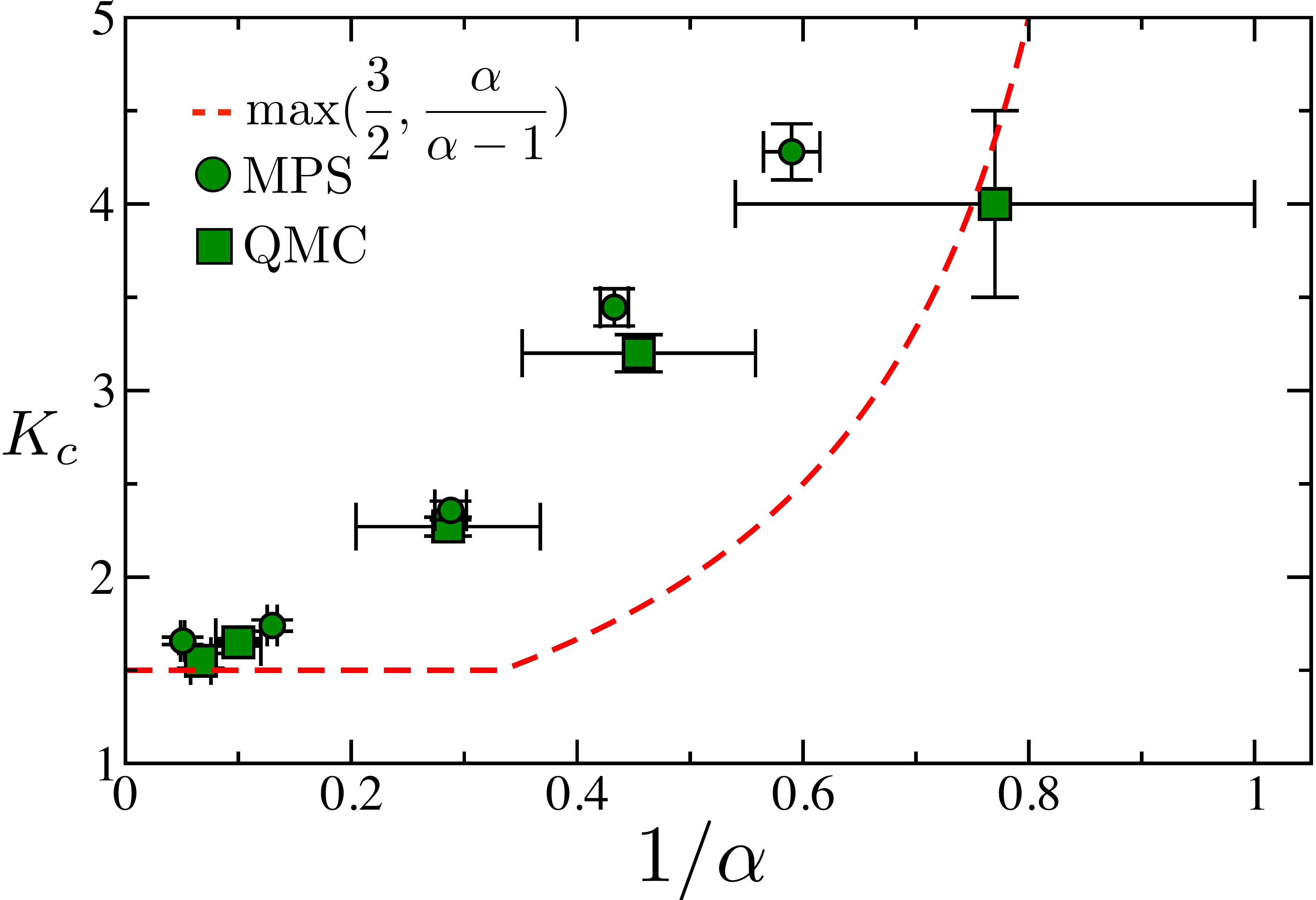}
 \caption{Critical LL parameter $K_c$ in the strong disorder regime plotted as a function of the inverse weak-link exponent $1/\alpha$. Both QMC and MPS estimates are shown together with the prediction from Yao {\it{et al.}}~\cite{Yao2016a} $K_c=\tilde{\alpha}/(\tilde{\alpha}-1)$ (dashed red line) where we assumed that the bare exponent $\tilde{\alpha}=\alpha$ due to scale invariance at criticality.}
 \label{fig:Yao}
\end{figure}

The prediction for the critical $\alpha$ can be compared to the prediction of Yao {\it et al.} \cite{Yao2016a}, assuming a direct correspondence between the bare power law coefficient $\tilde{\alpha}$ of the weak-link theory of Yao {\it et al.\ }and our numerical results for $\alpha$.
Such a direct correspondence is justified on the basis of the self-similarity of the distributions at criticality that we observe.
In Figure \ref{fig:Yao} we show the comparison between both predictions for the power law $\alpha$ at the critical LL parameter $K_c$.

\begin{figure}[!htb]
 \includegraphics[width=\columnwidth]{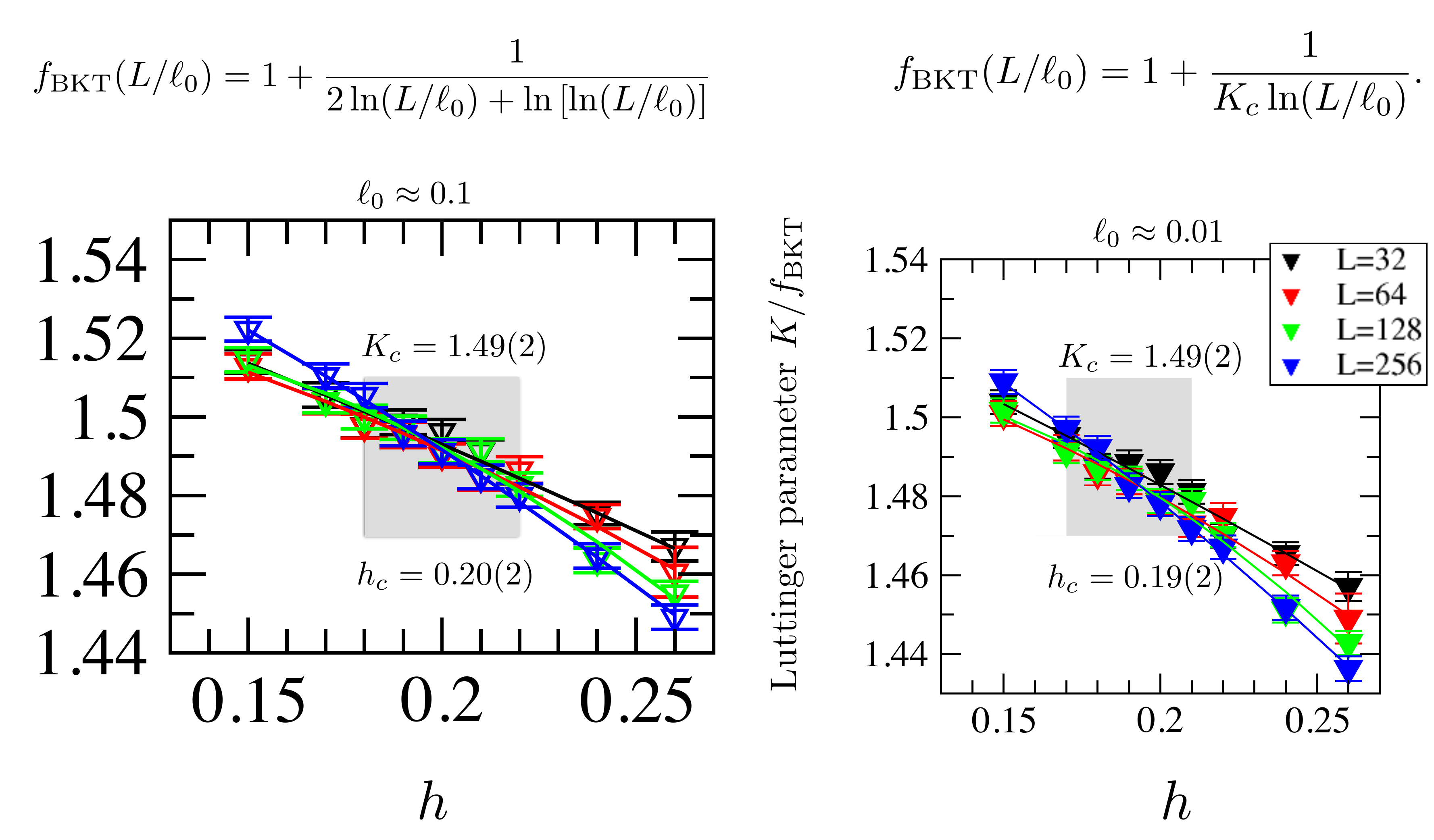}
 \label{fig:zoom}
 \caption{Crossing of the Luttinger parameter $K/f_{\rm BKT}$ for various system sizes using two slightly different finite size scaling forms, as indicated on top of both panels. QMC data for the XXZ chain at $\Delta=-0.55$. In both cases, the critical field is the same within error bars as well as the crossing point which gives the universal jump $K_c\simeq 1.5$ in excellent agreement with the Giamarchi-Schulz prediction.}
\end{figure}

\section{Finite size scaling of the superfluid density at criticality}
Here we want to discuss the logarithmic corrections at the SF-BG transition. While the standard BKT transition leads to the following finite size corrections~\cite{Weber1988a,Hsieh2013} $f_{\rm BKT}(L/\ell_0)=1+\frac{1}{2\ln(L/\ell_0)+\ln\left[\ln(L/\ell_0)\right]}$, the integration of the Giamarchi-Schulz renormalization equations leads at the lowest order to a slightly different form: $f_{\rm BKT}(L/\ell_0)=1+\frac{1}{K_c\ln(L/\ell_0)}$. 

In the main text (Fig. 2), we have therefore used the latter finite size expression to analyse our QMC data for the Luttinger parameter in the critical region. In Fig.~\ref{fig:zoom} we show two panels for our QMC data at $\Delta=-0.55$ using the aforementioned logarithmic corrections. Given our finite size samples, we have observed that both forms give a similarly good description, and the critical parameters ($h_c,K_c$) remain unchanged within error bars.

\bibliography{ref}
\bibliographystyle{apsrev4-1}

\end{document}